\RequirePackage{fixltx2e}
\documentclass[twocolumn,hyperpdf, amsmath,amssymb, aps,prl, reprint, superscriptaddress,nofootinbib,preprintnumbers]{revtex4-1}

\usepackage{graphicx}
\usepackage{dcolumn}
\usepackage{bm}
\usepackage{subfigure} 
\usepackage{amsmath}
\usepackage{amsfonts}
\usepackage{slashed}
\usepackage{hyperref}
\usepackage{multirow}
\usepackage{soul}
\usepackage{mathtools}
\usepackage{float}
\usepackage{afterpage}
\usepackage{color}
\usepackage{latexsym} 
\usepackage{rotating} 
\usepackage{verbatim} 
\usepackage{microtype} 
\usepackage{marginnote}
\usepackage[utf8]{inputenc}

\hypersetup{%
pdftitle = {Isoscalar $\pi\pi$ scattering and the $\sigma$ meson resonance from QCD},
pdfsubject = {QCD},
pdfkeywords = {QCD, Hadron, Physics, JLab, Lattice, Meson, Scattering},
pdfauthor = {Hadron Spectrum Collaboration},
colorlinks = {true},
filecolor = {black},
linkcolor = {black},
menucolor = {black},
citecolor = {blue},
urlcolor = {blue},
}{}

\begin{document}

\preprint{JLAB-THY-16-2287}
\preprint{DAMTP-2016-47}

\title{
Isoscalar $\pi\pi$ scattering and the $\sigma$ meson resonance from QCD
}


\author{Raul~A.~Brice\~{n}o}
\email{briceno@jlab.org}
\affiliation{Thomas Jefferson National Accelerator Facility, 12000 Jefferson Avenue, Newport News, VA 23606, USA}
\affiliation{Department of Physics, Old Dominion University, Norfolk, VA 23529, USA}

\author{Jozef~J.~Dudek}
\email{dudek@jlab.org}
\affiliation{Thomas Jefferson National Accelerator Facility, 12000 Jefferson Avenue, Newport News, VA 23606, USA}
\affiliation{Department of Physics, Old Dominion University, Norfolk, VA 23529, USA}

\author{Robert~G.~Edwards}
\email{edwards@jlab.org}
\affiliation{Thomas Jefferson National Accelerator Facility, 12000 Jefferson Avenue, Newport News, VA 23606, USA}

\author{David~J.~Wilson}
\email{d.j.wilson@damtp.cam.ac.uk}
\affiliation{Department of Applied Mathematics and Theoretical Physics, Centre for Mathematical Sciences, University of Cambridge, Wilberforce Road, Cambridge, CB3 0WA, UK}


\collaboration{for the Hadron Spectrum Collaboration}
\date{\today}

\begin{abstract}
We present for the first time a determination of the energy dependence of the isoscalar $\pi\pi$ elastic scattering phase-shift within a first-principles numerical lattice approach to QCD. Hadronic correlation functions are computed including all required quark propagation diagrams, and from these the discrete spectrum of states in the finite volume defined by the lattice boundary is extracted. From the volume dependence of the spectrum we obtain the $S$-wave phase-shift up to the $K\overline{K}$ threshold. Calculations are performed at two values of the $u,d$ quark mass corresponding to $m_\pi = 236, 391$ MeV and the resulting amplitudes are described in terms of a $\sigma$ meson which evolves from a bound-state below $\pi\pi$ threshold at the heavier quark mass, to a broad resonance at the lighter quark mass.
\end{abstract}

\maketitle


\emph{Introduction:}~Meson-meson scattering has long served as a tool to investigate the fundamental theory of strong interactions, quantum chromodynamics (QCD). The isoscalar channel, where all flavor quantum numbers are equal to zero, is dominated at low energies by $\pi\pi$ scattering, but despite experimental data on elastic $\pi\pi$ scattering being in place for many decades~\cite{Protopopescu:1973sh, *Hyams:1973zf, *Grayer:1974cr, *Estabrooks:1974vu}, the existence of the lowest lying resonance with spin zero, the $f_0(500)/\sigma$, has only recently been demonstrated with certainty~\cite{Caprini:2005zr, GarciaMartin:2011jx}. This is astonishing given the crucial role played by such a state in our understanding of, for example, spontaneous chiral symmetry breaking~\cite{GellMann:1960np}, and long-range contributions to the nuclear force~\cite{Johnson:1955zz}. The difficulty comes from the especially short lifetime of the $\sigma$ which causes it to lack the simple narrow ``bump'' signature associated with longer-lived resonances. It is the use of dispersive analysis techniques~\cite{Roy:1971tc}, which build in constraints from the causality and crossing symmetry of scattering amplitudes, when applied to the experimental data, which have led to an unambiguous signal for a $\sigma$ resonance. These techniques have ensured that the location of the corresponding pole singularity, located far into the complex energy plane, can now be stated with a high level of precision~\cite{Pelaez:2015qba, Agashe:2014kda}.

In principle it should be possible within QCD to calculate the scalar, isoscalar $\pi\pi$ scattering amplitude and the contribution of the $\sigma$ resonance to it, but the non-perturbative nature of the theory at low energies leaves us with limited calculational tools. The most powerful current approach is \emph{lattice QCD}, in which the quark and gluon fields are discretized on a space-time grid of finite size, allowing numerical computation by averaging over large numbers of possible field configurations generated by Monte-Carlo. In particular, from the time-dependence of correlation functions calculated in this way, we can extract a discrete spectrum of states whose dependence on the volume of the lattice can be related to meson-meson scattering amplitudes~\cite{Luscher:1990ux, *Rummukainen:1995vs, *Kim:2005gf, *Fu:2011xz, *Leskovec:2012gb, He:2005ey, *Hansen:2012tf, *Briceno:2012yi, *Guo:2012hv}.

Calculations of the scalar, isoscalar channel have long been considered to be among the most challenging applications of lattice QCD. In order to be successful here it is necessary to evaluate all quark propagation diagrams contributing to the correlation functions, to reliably extract a large number of states in the spectrum, and to determine and interpret the energy dependence of the scattering amplitude in the elastic region. To date no calculation has overcome all these challenges~\cite{Alford:2000mm, *Prelovsek:2010kg, *Fu:2013ffa, *Wakayama:2014gpa, *Howarth:2015caa, *Bai:2015nea}.

In this Letter we show that by combining a number of novel techniques whose application we have pioneered over the past few years, we can meet all these challenges and provide the first determinations of the scalar, isoscalar scattering amplitude within QCD. By utilizing \emph{distillation}~\cite{Peardon:2009gh}, we are able to evaluate with good statistical precision all required quark propagation diagrams, including those which feature quark-antiquark annihilation. By diagonalizing matrices of correlation functions~\cite{Michael:1985ne, *Dudek:2007wv, *Blossier:2009kd} using a large basis of composite QCD operators with relevant quantum numbers~\cite{Dudek:2009qf, Dudek:2010wm, Dudek:2010ew, Dudek:2011tt, Thomas:2011rh, Dudek:2012gj, Dudek:2012xn} we are able to make robust determinations of spectra, and by considering multiple lattice volumes and moving frames, we are able to map out the energy dependence of the $\pi\pi$ scattering amplitude over the entire elastic region.

We perform our calculations with two different values of the degenerate $u,d$ quark mass, corresponding to pions of mass 236 MeV and 391 MeV. We find that for the lighter mass, the scattering amplitude is compatible with featuring a $\sigma$ appearing as a broad resonance, which closely resembles the experimental situation. As the quark mass is increased we find that the $\sigma$ evolves into a stable bound-state lying below $\pi\pi$ threshold. 

\pagebreak

\emph{Correlation functions and the finite-volume spectrum:}~

The discrete spectrum of hadronic eigenstates of QCD in a finite volume is extracted from two-point correlation functions, 
${   C_{ab }(t,t';\vec{P})=\langle 0|\mathcal{O}^{}_a(t,\vec{P})\mathcal{O}^\dag_b(t',\vec{P})|0\rangle  }$, with spatial momentum $\vec{P}=\frac{2\pi}{L}\big[n_x,n_y,n_z\big],$ where  $n_i\in \mathbb Z$ in an $L \times L \times L$ box.
We use a large basis of interpolating fields, $\mathcal{O}_a$, from two classes. The first are ``single-meson''-like operators~\cite{Dudek:2009qf, Dudek:2010wm, Thomas:2011rh} which resemble a $q\bar{q}$ construction of definite momentum, $(\bar\psi\mathbf{\Gamma}\psi)_{\vec{P}}$, where $\mathbf{\Gamma}$ are operators acting in spin, color and position space~\cite{Peardon:2009gh}. Both $u\bar{u} + d\bar{d}$ and $s\bar{s}$ flavor constructions are included~\cite{Dudek:2011tt, Dudek:2013yja}. The second class of operators are those resembling a pair of pions, ``$\pi\pi$'', with definite relative and total momentum, 
${ \sum_{\hat{p}_1, \hat{p}_2} w_{\vec{p}_1, \vec{p}_2; \vec{P}} \,  (\bar\psi\mathbf{\Gamma_1}\psi)_{\vec{p}_1}  (\bar\psi\mathbf{\Gamma_2}\psi)_{\vec{p}_2}    }$
~\cite{Dudek:2012gj}, projected into isospin=0. Each isovector pion-like operator is constructed as the particular linear superposition, in a large basis of single-meson operators, that maximally overlaps with the pseudoscalar ground state~\cite{Thomas:2011rh, Dudek:2012gj}.  

We compute matrices of correlation functions $C_{ab}(t,t';\vec{P})$ using multiple single-meson operators along with several relative momentum constructions, ${\vec{p}_1+\vec{p}_2=\vec{P}}$, of the $\pi\pi$-like operators\footnote{We also include several ``$K\overline{K}$''-like operators, of analogous construction to the ``$\pi\pi$'' operators, although they are not vital in the determination of the spectrum below $K\overline{K}$-threshold.}. This kind of operator basis has been used successfully in the determination of scattering amplitudes in the $\pi\pi$ $I=1$ channel~\cite{Dudek:2012xn, Wilson:2015dqa} and the coupled-channel $(\pi K, \eta K)$~\cite{Dudek:2014qha, *Wilson:2014cna} and $(\pi \eta, K\overline{K})$~\cite{Dudek:2016cru} cases.

\begin{figure}[t]
\begin{center}
\includegraphics[width = 0.85\columnwidth]{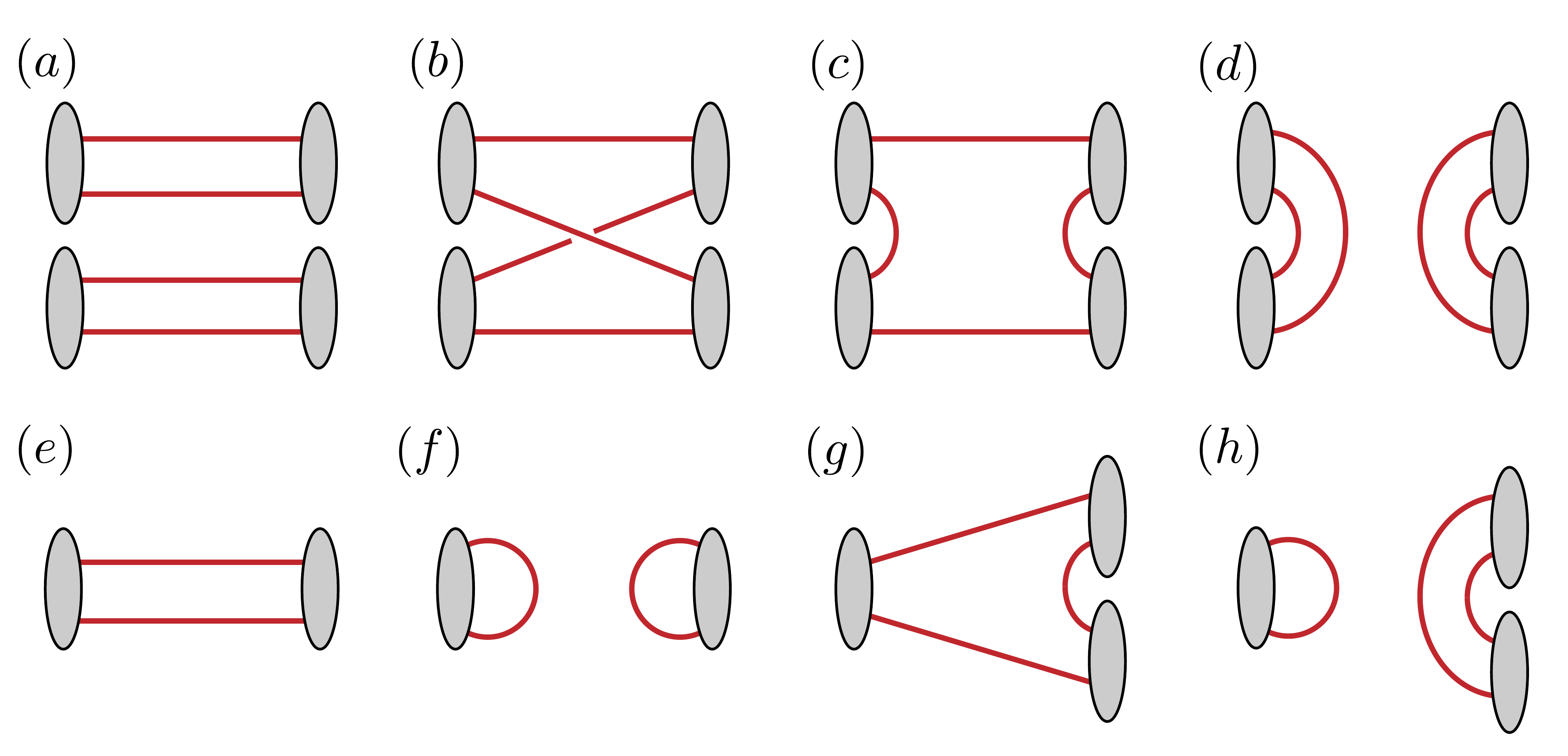}
\caption{Schematic quark propagation diagrams which contribute to the isoscalar correlation functions required in this Letter. 
}\label{fig:wicks}
\end{center}
\end{figure}

After integration over the quark fields appearing in the path-integral representation of $C_{ab}(t,t';\vec{P})$, we find that a variety of topologies of quark propagation diagrams appear, shown schematically in Fig.~\ref{fig:wicks}. Correlators with $\pi\pi$-like operators at $t$ and $t'$, for instance, require both \emph{connected} pieces $(a),(b)$ and partially $(c)$ and completely $(d)$ \emph{disconnected} pieces which feature quark propagation from a time $t$ to the same time $t$. Computation of these propagation objects has historically been a major challenge for lattice QCD. Within the distillation approach we utilize, determining these objects becomes manageable, and by obtaining them for all timeslices, $t$, good signals can be garnered by averaging correlation functions for fixed time separations over the whole temporal extent of the lattice. The factorization of operator construction, inherent in distillation, allows for the reuse of these propagation objects and those used here have been previously computed and used in other projects that featured quark annihilation~\cite{Dudek:2012xn, Dudek:2013yja, Wilson:2015dqa, Dudek:2014qha, Wilson:2014cna, Dudek:2016cru, Briceno:2015dca, Briceno:2016kkp}. 

\begin{figure}[t]
\begin{center}
\includegraphics[width = \columnwidth]{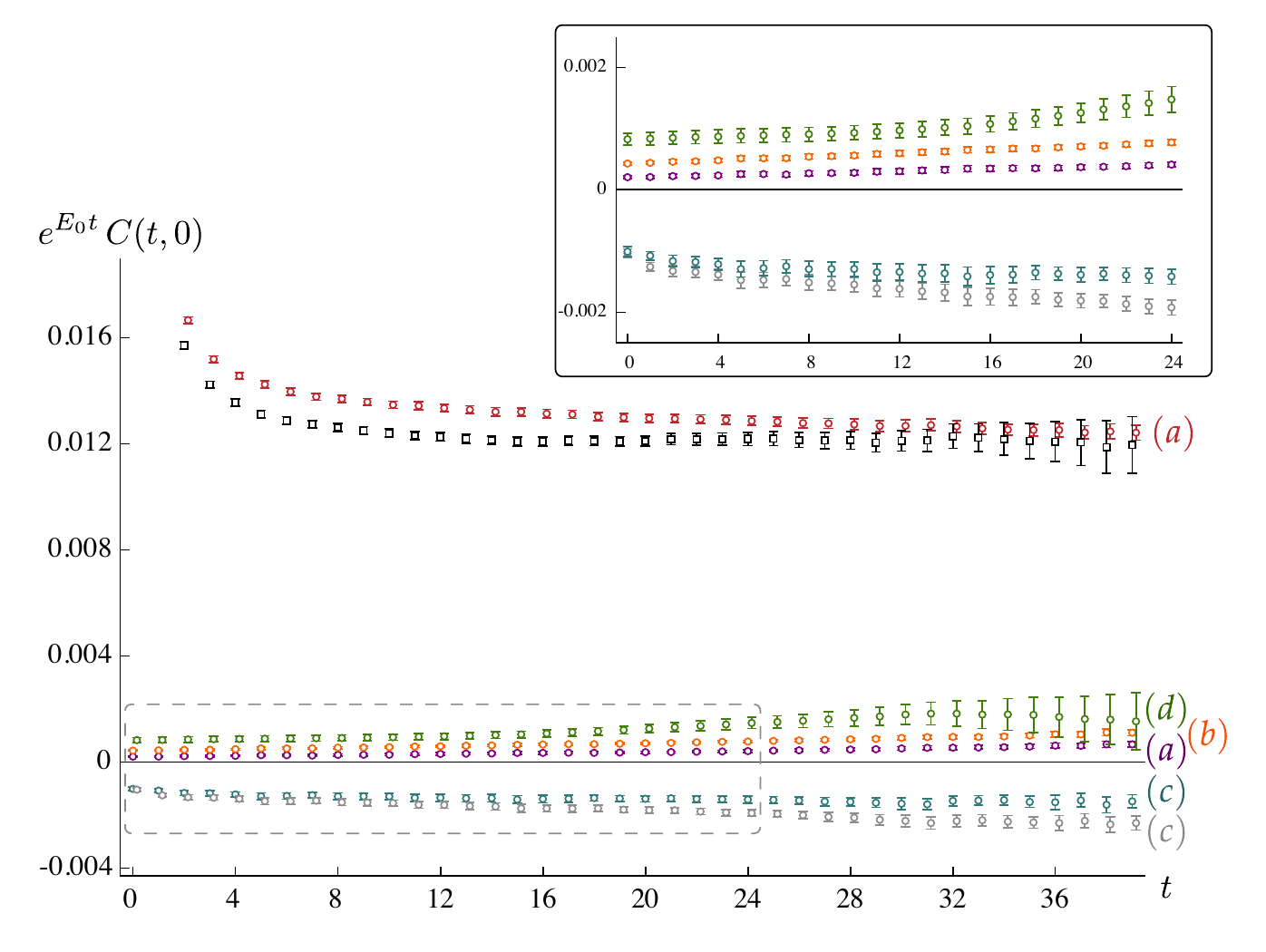}
\caption{Contributions of various diagrams (falling into the categories $(a)$--$(d)$ presented in Figure 1 -- two different variants of each of $(a)$ and $(c)$ appear) to the correlation function having an operator $\pi_{[000]} \pi_{[110]}$ at both $t'=0$ and $t$. The time dependence is weighted by $e^{E_0 t}$ with $E_0$ the energy of the lightest state with $\vec{P}=[110]$. The complete correlation function, which corresponds to the sum of the pieces shown, is shown by the black squares. Computation on a $32^3 \times 256$ lattice with $m_\pi = 236$ MeV.
}\label{fig:corr}
\end{center}
\end{figure}

In Fig.~\ref{fig:corr} we show the contributions of the various diagrams to an example correlation function having an operator $\pi_{[000]} \pi_{[110]}$ at both $t'=0$ and $t$, where we observe that all diagrams are evaluated with good statistical precision. In general delicate cancellations between different contributing diagrams can be present in isoscalar correlation functions, and our approach is seen to be capable of accurately capturing these.

\begin{figure}[t]
\begin{center}
\includegraphics[width = \columnwidth]{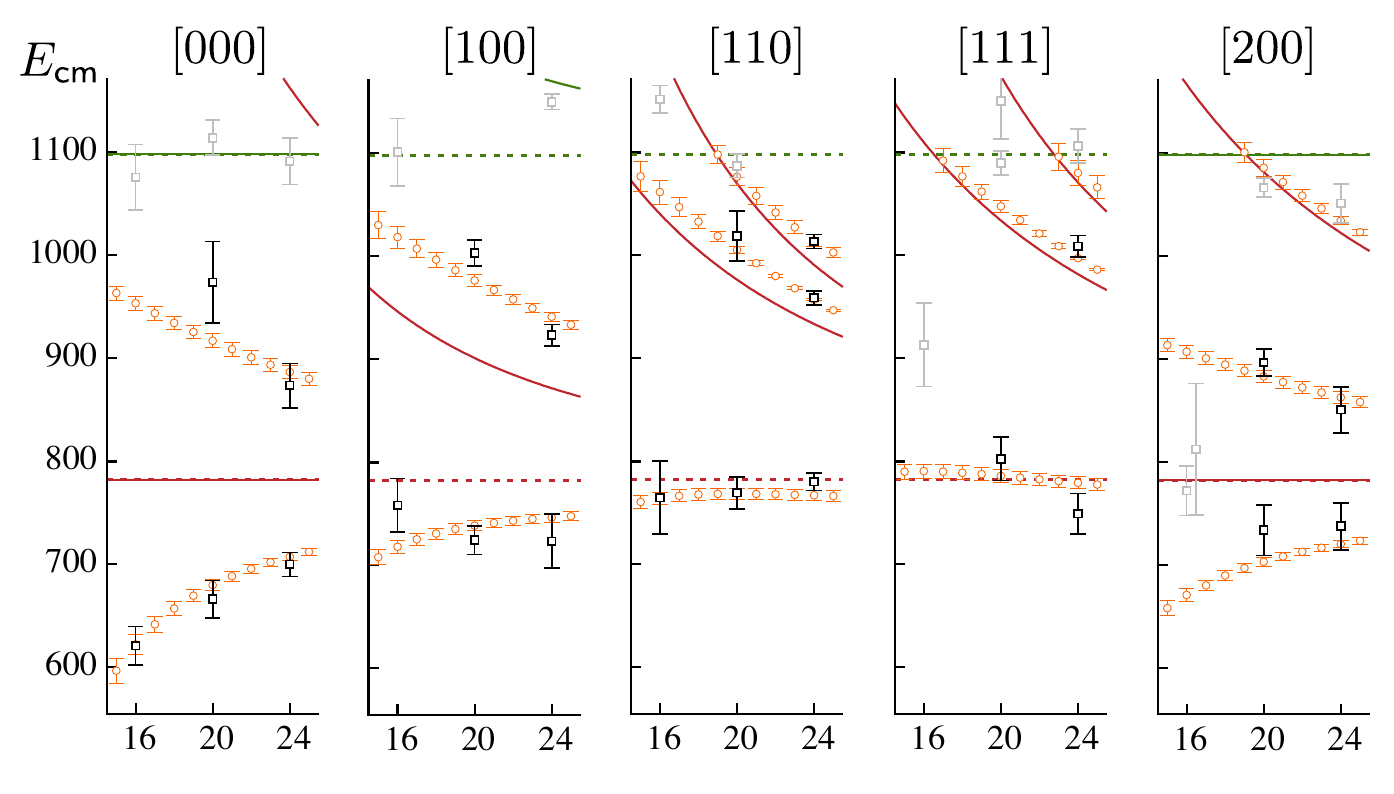}
\includegraphics[width = \columnwidth]{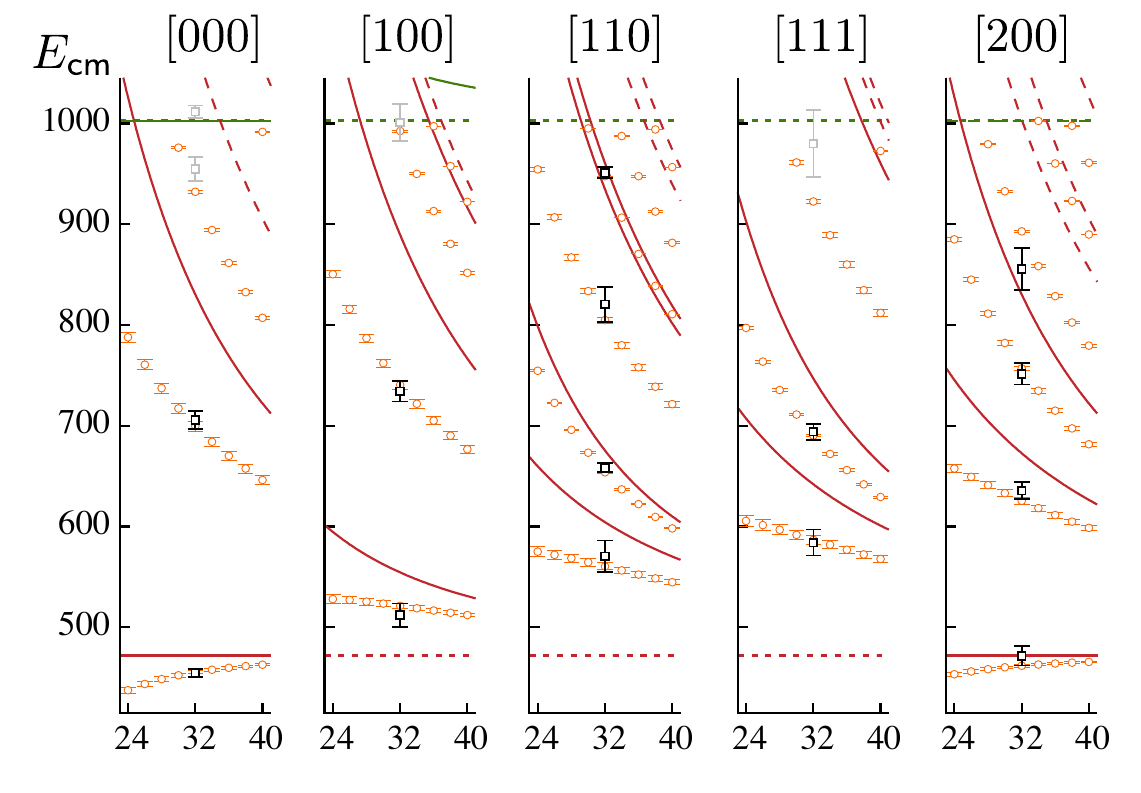}
\caption{Center-of-momentum frame energies (black and gray points) extracted from variational analysis of correlation matrices at five values of $\vec{P}$ plotted versus $L$. Upper panel shows spectra for $m_\pi = 391$ MeV and the lower panel for $m_\pi = 236$ MeV. The vacuum contribution to $[000]$ correlation functions, which is time-independent, and thermal contributions for other $\vec{P}$ are removed using the technique described in Ref.~\cite{Dudek:2012gj}. Red curves indicate non-interacting $\pi\pi$ energies. Small dashed red and green horizontal lines show the $\pi\pi$ and $K\overline{K}$ thresholds. Orange points show the finite-volume spectrum obtained from the $K$-matrix ``pole plus constant'' parameterization mentioned in the text.
}\label{fig:spec}
\end{center}
\end{figure}

We computed correlation matrices for total momentum $\vec{P} = [000], [100], [110], [111]$ and $[200]$, extracting multiple states in the spectrum of each using variational analysis of the type described in Ref.~\cite{Dudek:2010wm}. Details of the dynamical lattices, which include degenerate light $u,d$ quarks and a heavier $s$ quark, and which have spatial lattice spacing $a_s \sim 0.12 \, \mathrm{fm}$, can be found in Refs.~\cite{Lin:2008pr, Wilson:2015dqa}. For the 391 MeV pion case we computed with three lattice volumes, $16^3$, $20^3$ and $24^3$, while for the 236 MeV pion case we used a single larger $32^3$ volume. The extracted spectra are shown in Figure~\ref{fig:spec}. In this first study we will restrict our attention to energies below the $K\overline{K}$ threshold from which we can determine the $\pi\pi$ elastic scattering phase-shift.

\vspace{.5cm}

\emph{Scattering amplitudes and the $\sigma$-pole:}~ Under the well-justified approximation of neglecting kinematically suppressed higher partial waves, the $L\times L \times L$ finite-volume spectrum is related to the $S$-wave $\pi\pi$ elastic scattering phase-shift by
\begin{equation}
\cot \delta_0(E_\mathsf{cm}) + \cot \phi(P, L) = 0 \label{luscher}
\end{equation}
where $\phi(P,L)$ is a known function which differs according to $\vec{P}$~\cite{Luscher:1990ux, *Rummukainen:1995vs, *Kim:2005gf, *Fu:2011xz, *Leskovec:2012gb}. This provides a one-to-one mapping between the discrete finite-volume energies determined in lattice QCD and the infinite-volume scattering phase-shift evaluated at those energies. 

\begin{figure}[t]
\begin{center}
\includegraphics[width = \columnwidth]{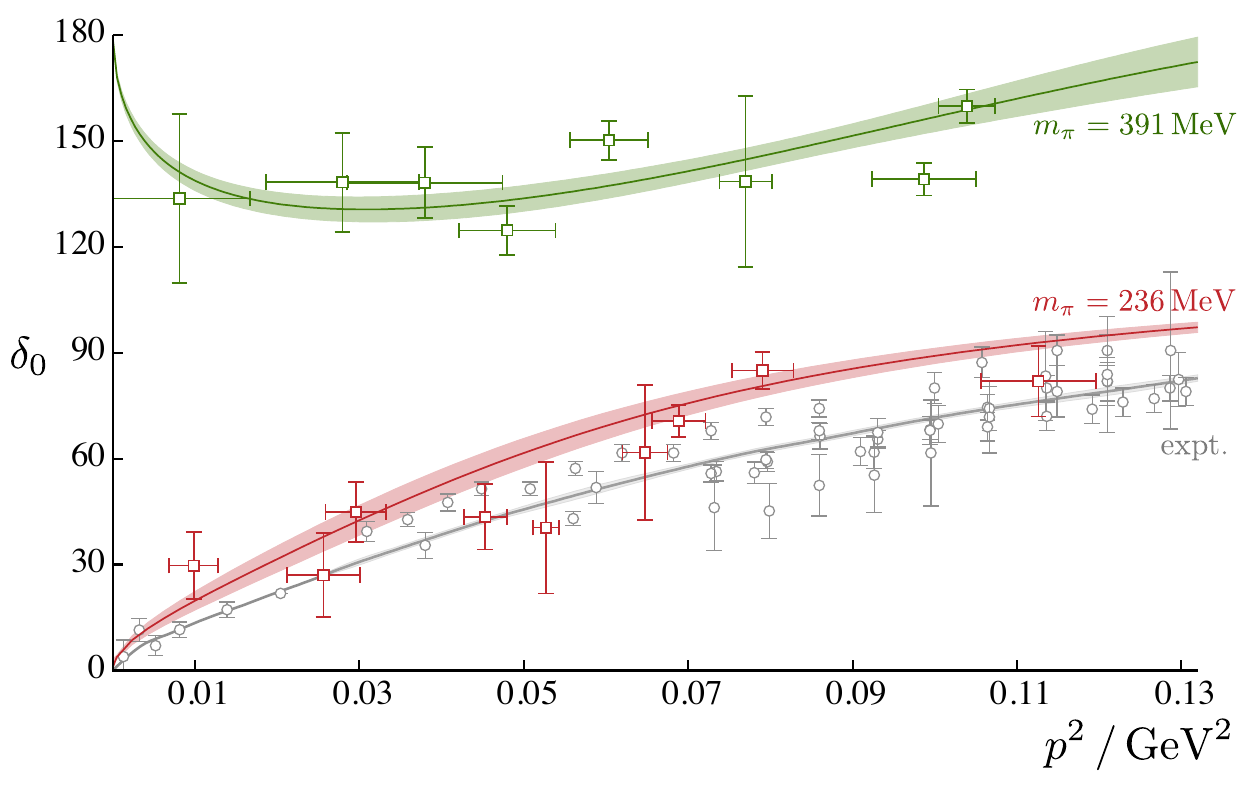}
\includegraphics[width = \columnwidth]{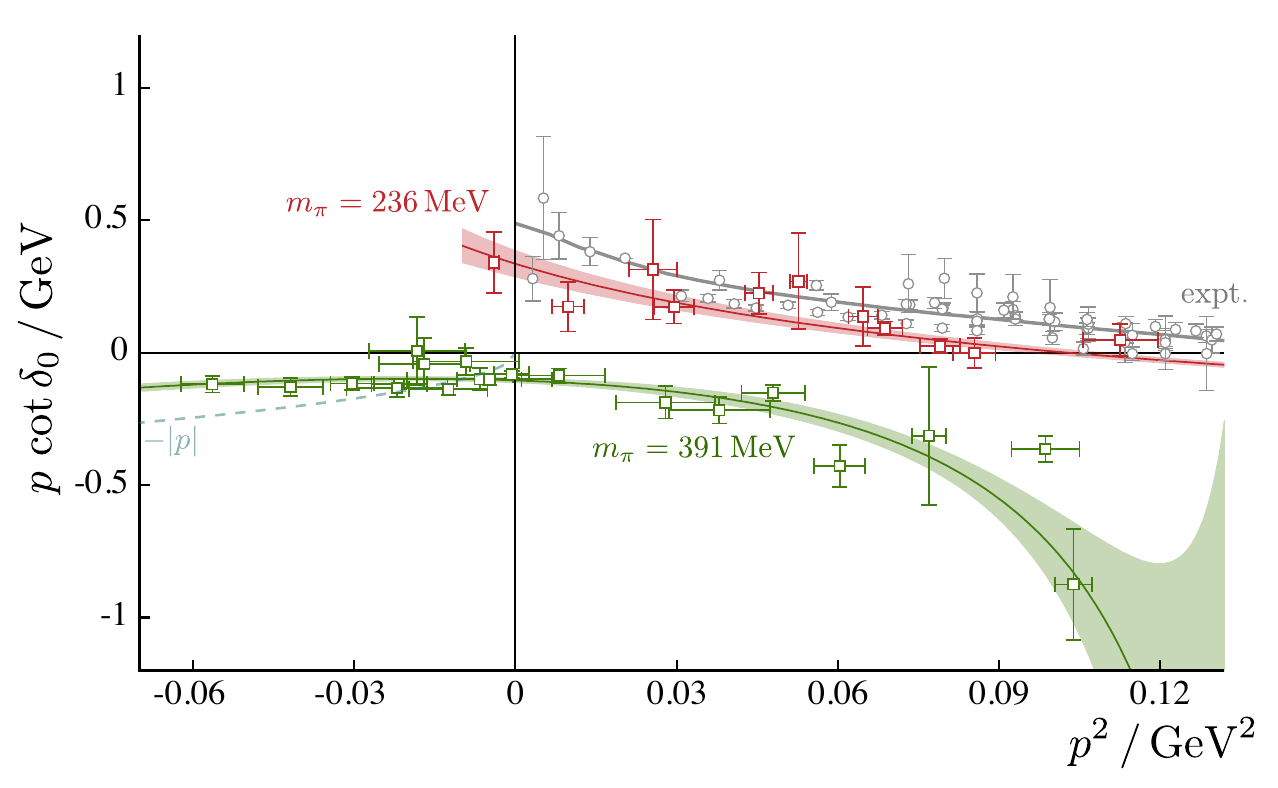}
\caption{Upper panel: $S$-wave $\pi\pi$ elastic scattering phase-shift, $\delta_0$, plotted against the scattering momentum, ${p^2 = ( E_\mathsf{cm}/2)^2 - m_\pi^2}$. Lower panel: Same data presented as $p \cot \delta_0$ (some points with large uncertainty have not been plotted). The colored curves are the result of a $K$-matrix ``pole plus constant'' with Chew-Mandelstam phase-space parameterization. The gray points show experimental data~\cite{Protopopescu:1973sh, *Hyams:1973zf, *Grayer:1974cr, *Estabrooks:1974vu} and the gray curve shows the constrained dispersive description of these data presented in Ref.~\cite{GarciaMartin:2011jx}.
}\label{fig:amps}
\end{center}
\end{figure}

In Figure~\ref{fig:amps} we present the phase-shifts determined from the spectra shown in Figure~\ref{fig:spec}. A simple-minded approach to parameterizing the energy dependence of these scattering amplitudes neglects the explicit contribution of any left-hand cut\footnote{which is present due to crossing symmetry, but which is distant from the physical scattering region for heavy pions.}, leaving significant freedom in choice of functional form. We find that we can obtain good descriptions of the lattice spectra for many unitarity-preserving choices of parameterization -- Figure~\ref{fig:amps} shows one illustrative example, which uses a single-channel $K$-matrix featuring a pole plus a constant, and a Chew-Mandelstam phase-space (see Ref.~\cite{Wilson:2014cna} and references therein), the corresponding description of the finite volume spectrum being shown in orange in Figure~\ref{fig:spec}. In previous studies~\cite{Dudek:2012xn, Wilson:2015dqa, Dudek:2014qha, Wilson:2014cna, Dudek:2016cru} of amplitudes featuring narrow resonances, we observed very little variation in the pole position of the resonance with parameterization choice variation. 

In the 391 MeV pion case, we find that all parameterizations which successfully describe the finite volume spectra have a pole on the real energy axis below $\pi\pi$ threshold on the physical Riemann sheet, which we interpret as the $\sigma$ appearing as a bound state of mass $758(4)$ MeV. Considering the amplitude determined with 236 MeV pions, we observe in Figure~\ref{fig:amps} a qualitative change of behavior in the phase-shift curve to a form which does not resemble either that expected for a bound-state or that of a narrow elastic resonance. We find that all successful descriptions of the spectrum have a pole on the second Riemann sheet with a large imaginary part, which we interpret as the $\sigma$ appearing as a broad resonance. Because the amplitude, determined from the finite-volume spectrum, is only constrained on the real energy axis, which is far from the pole position, there is a significant variation in the precise determination of the location of the pole under reasonable variations of the parameterization form\footnote{see Refs.~\cite{Wilson:2014cna, Dudek:2016cru} for the kinds of variation we consider.}. This is the same phenomenon that is observed when experimental $\pi\pi$ phase-shift data are used to fix parameters in amplitude models that do not build in dispersive constraints~\cite{Pelaez:2015qba}.

\begin{figure}[t]
\begin{center}
\includegraphics[width = \columnwidth]{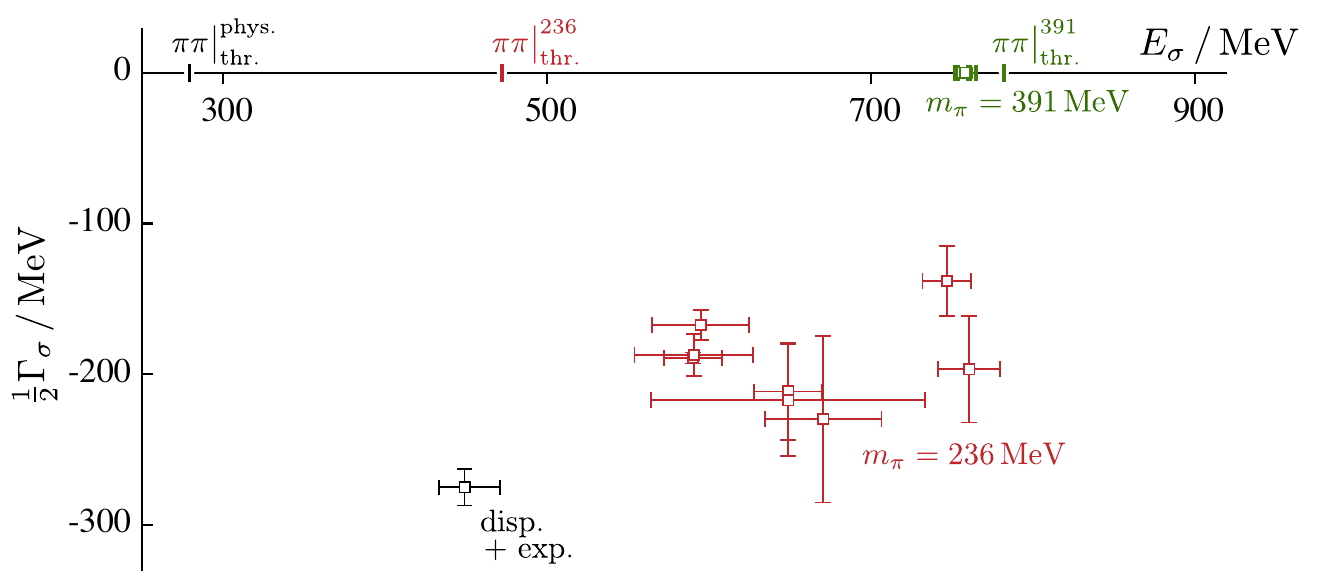}

\vspace{.4cm}

\includegraphics[width = \columnwidth]{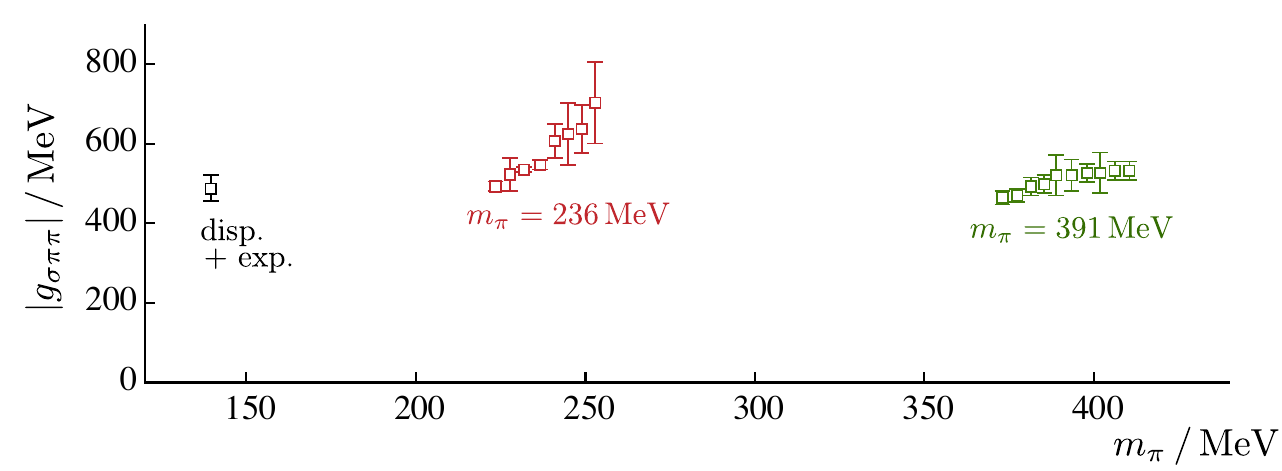}
\caption{Upper panel: $t$-matrix pole positions for a variety of parameterizations ($K$-matrix ``pole plus polynomial'' forms with and without Chew-Mandelstam phase-space and/or Adler zero~\cite{Adler:1964um}, relativistic Breit-Wigner and effective range expansion). Green points: bound-state pole (physical sheet) at $m_\pi = 391$ MeV. Red points: resonant pole (unphysical sheet) at $m_\pi = 236$ MeV. Black point: Resonant pole from dispersive analysis of experimental data (conservative average presented in Ref.~\cite{Pelaez:2015qba}). Lower panel: Coupling $g_{\sigma\pi\pi}$ from $t$-matrix residue at the pole - points from various parameterizations are shifted horizontally for clarity.
}\label{fig:poles}
\end{center}
\end{figure}

Figure~\ref{fig:poles} shows the complex energy plane illustrating the extracted pole position, $s_0 = (E_\sigma - \tfrac{i}{2} \Gamma_\sigma)^2$, for a range of parameterization choices. We also show the coupling, $|g_{\sigma \pi\pi}|$, extracted from the residue of the $t$-matrix at the pole, $g_{\sigma \pi\pi}^2 = \lim_{s\to s_0} (s_0 -s) \, t(s)$.

\vspace{0.5cm}


\emph{Summary and outlook:}~In this Letter we have, for the first time, determined the low-lying spectra of the scalar-isoscalar channel of QCD in a box, including all required quark propagation diagrams. From the finite-volume spectra we have extracted the $\pi\pi$ elastic scattering amplitude which shows qualitatively different behavior at the two pion masses considered, 236 MeV and 391 MeV, with the heavier mass featuring a $\sigma$ appearing as a stable bound-state.

The amplitude parameterizations we explored to describe the finite-volume spectrum determined with 236 MeV pions all feature a $\sigma$ appearing as a broad resonance, but the pole position is not precisely determined, showing variation with parameterization choice. We believe that this comes about because our parameterizations, while maintaining elastic unitarity, do not necessarily respect the analytical constraints placed on them by causality and crossing symmetry. In the future we plan to adapt dispersive approaches so that they are applicable to describing the lattice data, and we expect this will allow us to pin down the $\sigma$ pole position with precision directly from QCD.

With constrained amplitude forms in hand, it will become appropriate to perform calculations with lighter $u,d$ quarks, such that we move closer to the physical pion mass, in order to make direct comparison with the experimental situation. It will also be useful to examine pion masses between the 236 MeV and 391 MeV considered here to determine how the transition we have observed from bound-state to resonance is manifested -- a suggestion from unitarized chiral perturbation theory~\cite{Hanhart:2008mx, *Nebreda:2010wv} has the coupling $g_{\sigma \pi \pi}$, which one might conclude from Figure~\ref{fig:poles} is approximately independent of quark mass, having a divergent behavior somewhere near $m_\pi \sim 300$ MeV.

Our calculational techniques allow us to determine finite-volume spectra above the $K\overline{K}$ threshold, and by considering such energies within a coupled-channel analysis, we expect to be able to study any $f_0(980)$-like resonance that may appear. Such a state is anticipated as an isospin partner of the $a_0$ resonance which we observed near the $K\overline{K}$ threshold in a recent 391 MeV pion mass calculation~\cite{Dudek:2016cru}. A comprehensive study of the light scalar meson nonet ($\sigma, \kappa, a_0, f_0$) within first-principles QCD will then be possible. The finite-volume approach can also be extended to study the coupling of these states to external currents~\cite{Briceno:2014uqa,Briceno:2015csa,Briceno:2015dca,Briceno:2016kkp,Briceno:2015tza, Bernard:2012bi, Agadjanov:2014kha, Lellouch:2000pv} -- by examining the current virtuality dependence of the form-factors evaluated at the resonance pole, we expect to be able to infer details of the constituent structure of the scalar mesons.

\pagebreak
\emph{Acknowledgments:}
We thank our colleagues within the Hadron Spectrum Collaboration, and in particular, thank B\'alint Jo\'o for help. We also thank Kate Clark for use of the {\tt QUDA} codes. RAB would like to thank I.~Danilkin for useful discussions in the preparation of the manuscript.
The software codes
{\tt Chroma}~\cite{Edwards:2004sx}, {\tt QUDA}~\cite{Clark:2009wm,Babich:2010mu}, {\tt QUDA-MG}~\cite{Clark:SC2016}, {\tt QPhiX}~\cite{ISC13Phi}, and {\tt QOPQDP}~\cite{Osborn:2010mb,Babich:2010qb} were used for the computation of the quark propagators. The contractions were performed on clusters at Jefferson Lab under the USQCD Initiative and the LQCD ARRA project. This research was supported in part under an ALCC award, and used resources of the Oak Ridge Leadership Computing Facility at the Oak Ridge National Laboratory, which is supported by the Office of Science of the U.S. Department of Energy under Contract No. DE-AC05-00OR22725.
This research is also part of the Blue Waters sustained-petascale computing project, which is supported by the National Science Foundation (awards OCI-0725070 and ACI-1238993) and the state of Illinois. Blue Waters is a joint effort of the University of Illinois at Urbana-Champaign and its National Center for Supercomputing Applications. This research used resources of the National Energy Research Scientific Computing Center (NERSC), a DOE Office of Science User Facility supported by the Office of Science of the U.S. Department of Energy under Contract No. DE-AC02-05CH11231.
The authors acknowledge the Texas Advanced Computing Center (TACC) at The University of Texas at Austin for providing HPC resources. Gauge configurations were generated using resources awarded from the U.S. Department of Energy INCITE program at Oak Ridge National Lab, and also resources awarded at NERSC. RAB, RGE and JJD acknowledge support from U.S. Department of Energy contract DE-AC05-06OR23177, under which Jefferson Science Associates, LLC, manages and operates Jefferson Lab. JJD acknowledges support from the U.S. Department of Energy Early Career award contract DE-SC0006765. DJW acknowledges support from the Isaac Newton Trust/University of Cambridge Early Career Support Scheme [RG74916].


\bibliographystyle{apsrev4-1}
\bibliography{shortbib}

\end{document}